\documentclass[
 reprint,
 amsmath,amssymb,
 aps,nofootinbib,superscriptaddress
]{revtex4-2}

\usepackage{graphicx}
\usepackage{dcolumn}
\usepackage{bm}
\usepackage{xcolor}
\usepackage{float}
\begin{document}

\title{Charge doping into spin minority states mediates doubling of $T_\mathrm{C}$ in ferromagnetic CrGeTe$_3$}

\author{Liam Trzaska}
\affiliation{SUPA, School of Physics and Astronomy, University of St Andrews, St Andrews KY16 9SS, UK}

\author{Lei Qiao}
\affiliation{Physics Department, International Center of Quantum and Molecular Structures, Materials Genome Institute, State Key Laboratory of Advanced Special Steel, Shanghai Key Laboratory of High Temperature Superconductors, Shanghai University, Shanghai 200444, China}
\affiliation{Consiglio Nazionale delle Ricerche (CNR-SPIN), Unità di Ricerca presso Terzi c/o Università “G. D'Annunzio”, 66100 Chieti, Italy}

\author{Matthew D. Watson}
\affiliation{Diamond Light Source, Harwell Science and Innovation Campus, Didcot, OX11 ODE, United Kingdom}

\author{Monica \surname{Ciomaga Hatnean}}
\altaffiliation{Current address: Paul Scherrer Institut, Forschungsstrasse 111, 5232 Villigen PSI, Switzerland}
\affiliation {Department of Physics, University of Warwick, Coventry, CV4 7AL, United Kingdom}

\author{Igor Markovi\'{c}}
\author{Edgar Abarca Morales}
\affiliation{SUPA, School of Physics and Astronomy, University of St Andrews, St Andrews KY16 9SS, UK}
\affiliation{Max Planck Institute for Chemical Physics of Solids, N\"othnitzer Strasse 40, 01187 Dresden, Germany}
\author{Tommaso Antonelli}
\affiliation{SUPA, School of Physics and Astronomy, University of St Andrews, St Andrews KY16 9SS, UK}

\author{Cephise Cacho}
\affiliation{Diamond Light Source, Harwell Science and Innovation Campus, Didcot, OX11 ODE, United Kingdom}

\author{Geetha Balakrishnan}
\affiliation {Department of Physics, University of Warwick, Coventry, CV4 7AL, United Kingdom}

\author{Wei Ren}
\email{renwei@shu.edu.cn}
\affiliation{Physics Department, International Center of Quantum and Molecular Structures, Materials Genome Institute, State Key Laboratory of Advanced Special Steel, Shanghai Key Laboratory of High Temperature Superconductors, Shanghai University, Shanghai 200444, China}

\author{Silvia Picozzi}
\email{silvia.picozzi@spin.cnr.it}
\affiliation{Consiglio Nazionale delle Ricerche (CNR-SPIN), Unità di Ricerca presso Terzi c/o Università “G. D'Annunzio”, 66100 Chieti, Italy}

\author{Phil D.C. King}
\email{pdk6@st-andrews.ac.uk}
\affiliation{SUPA, School of Physics and Astronomy, University of St Andrews, St Andrews KY16 9SS, UK}

\date{\today}

\begin{abstract}
The recent discovery of the persistence of long-range magnetic order when van der Waals layered magnets are thinned towards the monolayer limit has provided a tunable platform for the engineering of novel magnetic structures and devices.  Here, we study the evolution of the electronic structure of CrGeTe$_3$ as a function of electron doping in the surface layer. From angle-resolved photoemission spectroscopy, we observe spectroscopic fingerprints that this electron doping drives a marked increase in $T_\mathrm{C}$, reaching values more than double that of the undoped material, in agreement with recent studies using electrostatic gating. Together with density functional theory calculations and Monte Carlo simulations, we show that, surprisingly, the increased $T_\mathrm{C}$ is mediated by the population of spin-minority Cr $t_{2g}$ states, forming a half-metallic 2D electron gas at the surface. We show how this promotes a novel variant of double exchange, and unlocks a significant influence of the Ge -- which was previously thought to be electronically inert in this system -- in mediating Cr-Cr exchange. \newline

\end{abstract}

\maketitle

\section{Introduction}
Dimensionality has a profound impact on the physical properties of materials. Control of the effective system dimensionality can be used to tune band gaps in semiconductors~\cite{chaves_2020},
control catalytic activity~\cite{Cao_2020}, and manipulate the strength of electronic interactions~\cite{king_atomic-scale_2014}. As such, it is of core importance both for fundamental understanding and technological application. This is true no more so than in the field of magnetism. While three-dimensional magnets are commonplace, long-range order is strictly forbidden to occur in one-dimensional systems~\cite{peierls_isings_1936}. Layered materials hosting magnetic ions present a novel environment in which to study the critical dimensionality between these two extremes~\cite{gibertini_magnetic_2019}: finite inter-plane coupling allows long-range order to develop, while the quasi-two-dimensionality of the system can be expected to have a strong influence on the magnetic anisotropy and the role of fluctuations.

Here, we study the quasi-2D layered magnet CrGeTe$_3$. In this compound, magnetic layers are separated by van der Waals gaps, leading to weak inter-layer interactions~\cite{Carteaux_1995,Ji2013,Liu_Critical_behavior}. Nonetheless, recent studies employing electrostatic gating have reported both a dramatic increase in the magnetic ordering temperature, $T_\mathrm{C}$, and a striking switch in the magnetic anisotropy with increasing gate voltage~\cite{Verzhbitskiy_CGT}, pointing to a subtle interplay between dimensionality and doping effects. Here, we utilise alkali-metal surface doping in ultra-high vacuum to mimic the conditions of field-effect doping, and probe the resulting changes in the surface electronic structure by angle-resolved photoemission spectroscopy (ARPES). Combined with density-functional theory (DFT) calculations, we demonstrate how electrostatic-type doping in this system drives the formation of a half-metallic 2D electron gas of minority-spin carriers, and show how this opens new exchange pathways which underpin the increase in $T_\mathrm{C}$.

\begin{figure*}
    \centering
    \includegraphics[width=\textwidth]{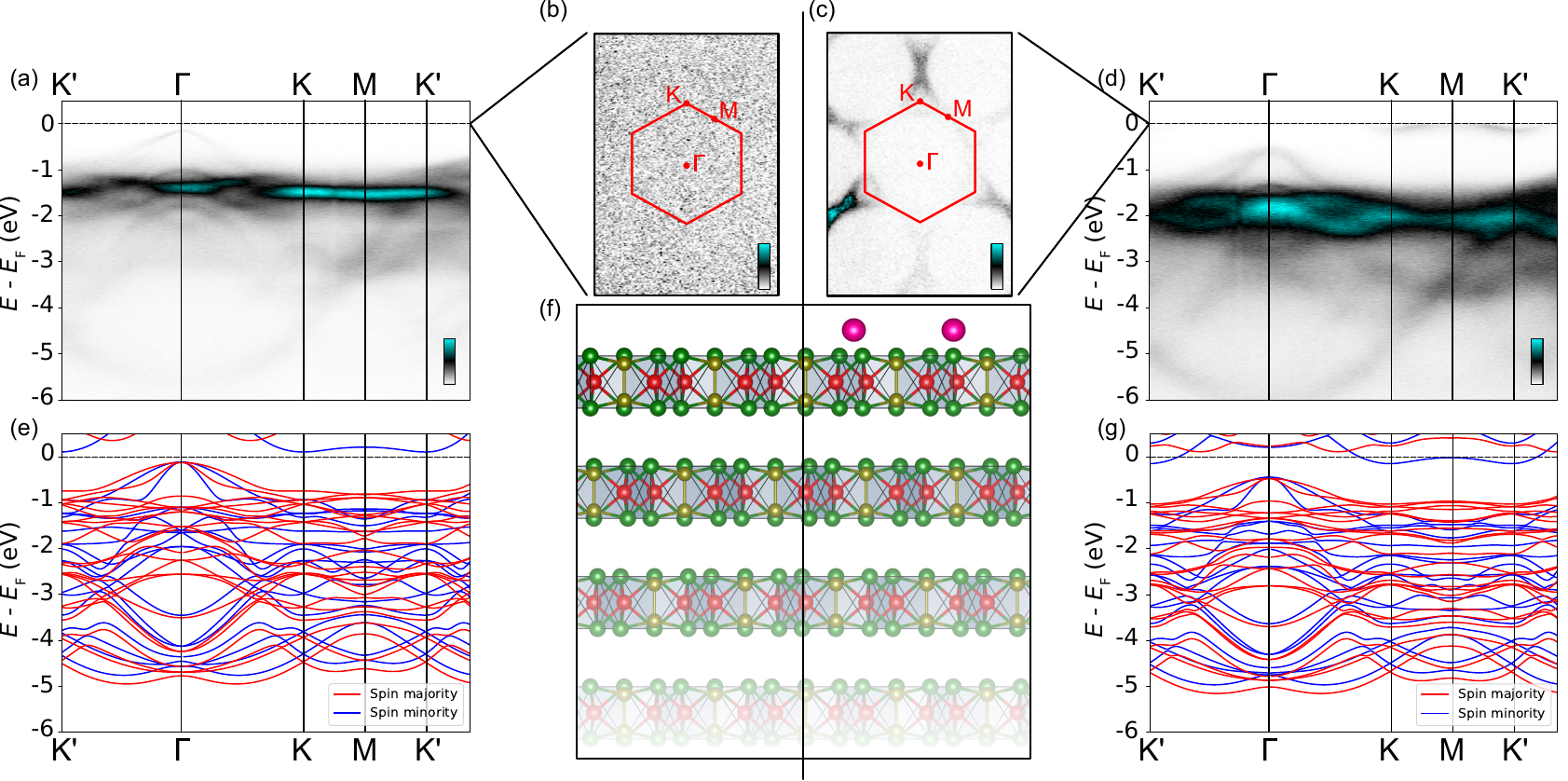}
    \caption{(a) ARPES dispersions measured ($h\nu=80$~eV) in the magnetic ground state along the $\rm K'-\Gamma-\rm K-\rm M-\rm K'$ direction of the Brillouin zone for undoped CrGeTe$_3$, reproduced from Ref.~\cite{Watson_CGT}. (b) Corresponding Fermi surface map. Consistent with the semiconducting nature of CrGeTe$_3$, we find no states at the Fermi level. (c,d) Equivalent to (b,a) for the Rb-doped surface. A clear set of Fermi pockets arises around the K points of the surface Brillouin zone. DFT band structure calculations for (e) undoped and (g) surface-doped CrGeTe$_3$. (f) Crystal structure of CrGeTe$_3$, showing a schematic of the Rb surface doping.}
    \label{Fig1}
\end{figure*}

\section{Results}
Figure~\ref{Fig1} shows an overview of the electronic structure of pristine CrGeTe$_3$, and the measured electronic structure following the deposition of sub-monolayer coverage of Rb atoms at the surface. In agreement with our DFT calculations (Fig.~\ref{Fig1}(e)), ARPES of the pristine compound (Fig.~\ref{Fig1}(a,b)) indicates several dispersive states near the top of the valence band, which are dominantly comprised of Te orbital character (see also Supplementary Fig.~1 for orbitally-resolved calculations). These approach, but do not reach, the Fermi level, with negligible spectral weight at $E_\mathrm{F}$ consistent with the semiconducting nature of this compound. The rather flat band observed at $E-E_\mathrm{F}\approx1-2$~eV is evident in our calculations as the spin-majority Cr-derived states. Consistent with previous calculations~\cite{wang_nikolaev_ren_solovyev_2019}, we find very little weight of the Ge states in the valence band here (see Supplementary Fig.~1). While Ge is crucial to obtain the chemically stable Cr$^{3+}$ charge state in CrGeTe$_3$, it can be considered as electrically inert in the parent compound. 

\begin{figure*}
    \centering
    \includegraphics[width=\textwidth]{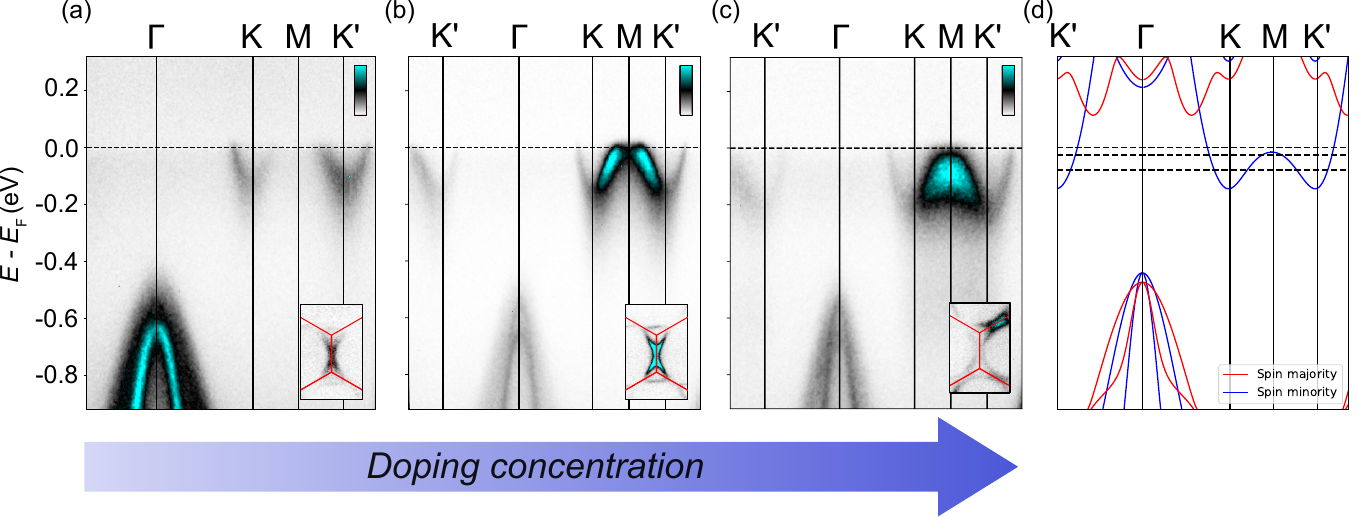}
    \caption{(a-c) ARPES dispersions measured along the $\Gamma-\rm K-\rm M-\rm K'$ direction of the Brillouin zone for a surface carrier density of (a) $N=3.1\times 10^{13}~\rm cm^{-2}$ ($h\nu=48$~eV), (b) $N=4.8\times 10^{13}~\rm cm^{-2}$ ($h\nu=80$~eV), and (c) $N=7.9\times 10^{13}~\rm cm^{-2}$ ($h\nu=80$~eV). Fermi surface measurements around the M point of the Brillouin zone are shown as insets. The topology of the Fermi surface changes, indicating a doping-dependent Lifshitz transition. (d) DFT calculated band structure, indicating the spin-minority nature of the conduction band. Calculations were performed with 0.2 electrons per Cr site. Dashed lines represent $E_{\rm F}$ for the ARPES measurements shown in (a-c).}
    \label{Fig2}
\end{figure*}

Upon surface Rb deposition (Fig.~\ref{Fig1}(d)), the dispersive Te-derived states and the flatter Cr-derived bands are both shifted downwards by around 500~meV (\ref{Fig1}(d)). This points to an efficient electron doping at the surface, where the deposited Rb atoms become readily ionised, donating their electrons into the near-surface region. Consistent with this, we find a finite population of the conduction band states around the K-M-K' path of the surface Brillouin zone. Our measurements indicate the conduction band minimum to be located at the K points (see also Fig.~\ref{Fig2}), indicating an indirect band gap in this system. Small triangular electron-like Fermi pockets are created around these zone-corner points, clearly visible in our Fermi surface measurements in Fig.~\ref{Fig1}(c). From photon energy-dependent measurements (Supplementary Fig. 2), we find that these states are non-dispersive along the out-of-plane direction, and are thus two-dimensional. We therefore conclude that the alkali metal dosing here leads to a near-surface band bending which, in turn, drives the formation of a 2D electron gas localised at the sample surface~\cite{Riley_WSe2}.

Interestingly, our DFT calculations for a Rb-doped CrGeTe$_3$ layer show that the lowest-energy conduction band, which becomes populated upon carrier doping (Fig.~\ref{Fig1}(g)), is of spin-minority character. In fact, previous calculations have come to different conclusions regarding the nature of the lowest conduction band in CrGeTe$_3$, and we find that the exact band ordering is sensitive to the value of the onsite energy $U$ used for the calculations (see Supplementary Fig.~3). Only the spin-minority band has the correct momentum-space dispersion to reproduce our experimental measurements, with band minima located at the zone-corner K points. We thus conclude that the low-$U$ regime provides the most accurate calculation scheme for the electron-doped system here, leading to the formation on doping of a two-dimensional half-metal of spin-minority character.

Fig.~\ref{Fig2} shows how this half-metallic 2DEG becomes populated with increasing surface carrier doping. At a surface doping density of $3.
1\times10^{13}$~cm$^{-2}$ (0.12 electrons/Cr, Fig.~\ref{Fig2}(a)), nearly V-shaped conduction bands are observed centred at the K and K' points of the Brillouin zone, with the conduction band crossing through the Fermi level along the $\Gamma$-K and K-M-K' lines. The resulting Fermi surface (Fig.~\ref{Fig2}(a), inset) is comprised of closed triangular pockets centred at the K and K' points. With increasing deposition of Rb on the surface, the conduction band pockets become more occupied, with a concomitant increase in the size of the triangular Fermi pockets (Fig.~\ref{Fig2}(b)). At a critical electron doping of around $5\times10^{13}$~cm$^{-2}$ (0.2 electrons/Cr), the Fermi level moves above a saddle point located at the M point of the Brillouin zone (Fig.~\ref{Fig2}(c)). This drives a Lifshitz transition~\cite{lifshitz}, where the triangular electron-like Fermi surfaces at low doping merge to form a large hole-like pocket centred at the Brillouin zone centre. These features are well described by our DFT calculations of the electronic structure of a monolayer of CrGeTe$_3$ (Fig.~\ref{Fig2}(d), see Methods, and also Supplementary Fig.~S4), indicating that the quantum-confined 2DEG here is well represented within a monolayer description.

\begin{figure*}
    \centering
    \includegraphics[width=\textwidth]{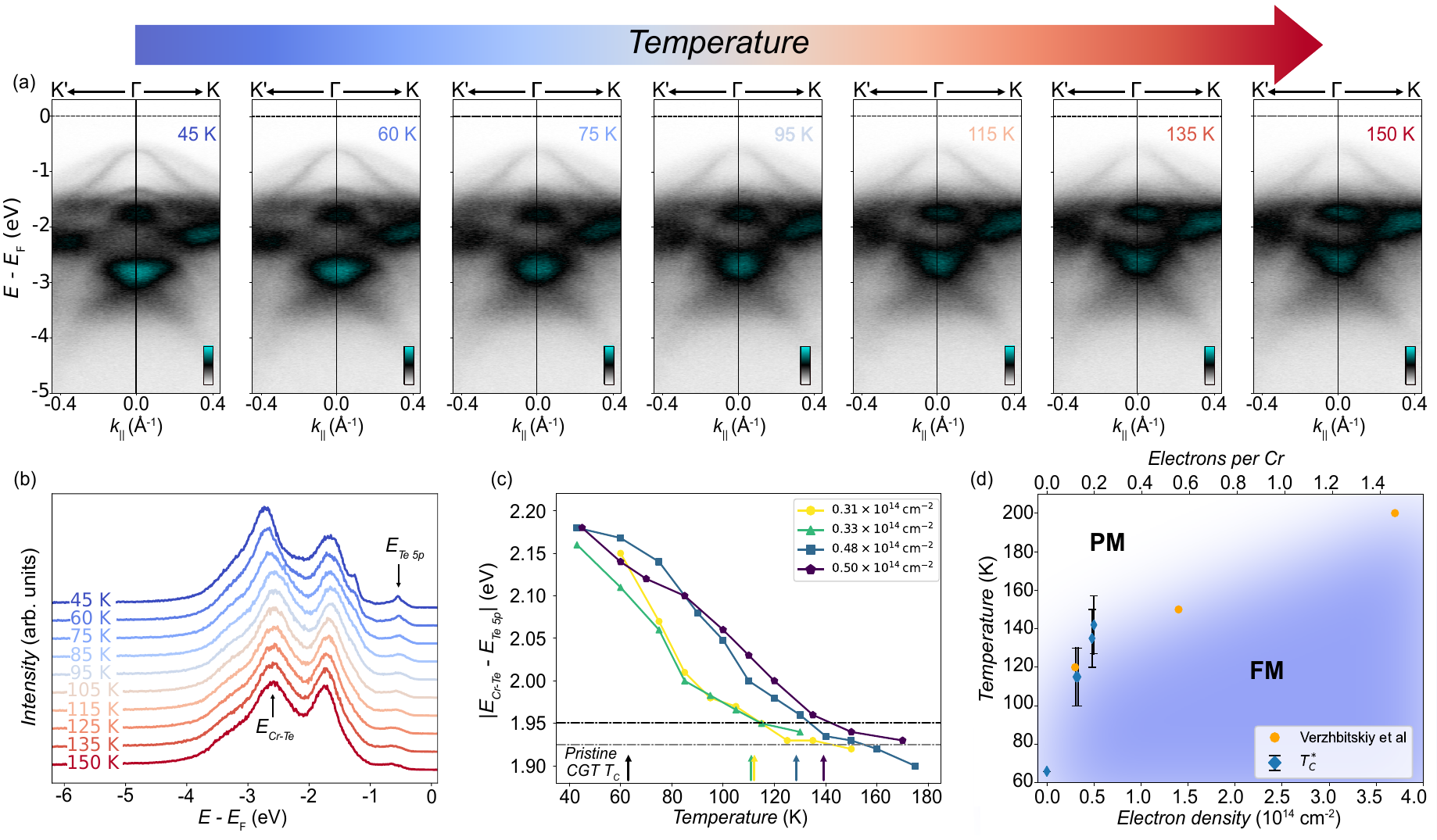}
    \caption{(a) ARPES measurements around the $\Gamma$ point ($h\nu=48eV$) for a CrGeTe$_3$ sample with carrier density $N=3.1\times 10^{13}~\rm cm^{-2}$, as a function of temperature. (b) Corresponding temperature-dependant EDCs extracted at $\Gamma$. (c) Extracted temperature-dependent separation of the Cr-Te hybridised state from the Te 5p valence band maximum (see arrows in (b)). Grey and black dashed lines represent an estimated baseline and a 10\% rise from that baseline. Approximate values for $T_{\rm C}$ are given by the intersection of each data set with the black dashed line, indicated by the coloured arrows. The magnetic $T_{\rm C}$ for pristine CrGeTe$_3$ is also shown. (d) Experimental phase diagram, plotting the extracted $T_{\rm C}$ estimates as a function of doping along with estimates from transport measurements from Ref.~\cite{Verzhbitskiy_CGT}.}
    \label{Fig3}
\end{figure*}
We now turn to the magnetic order of this surface-doped layer. Previously, some of us have identified a particular state in the valence band of the undoped compound -- derived from hybridised Te 5$p$ and Cr $e_g$ states -- which shows a strong temperature-dependent shift associated with the development of long-range magnetic order~\cite{Watson_CGT}. This band feature is evident also in our doped samples, for example in Fig.~\ref{Fig3}(a) as the bright spectral weight at the $\Gamma$ point at $E-E_\mathrm{F}\approx-2.9$~eV. As is evident in the raw data in Fig.~\ref{Fig3}(a), and in energy distribution curves (EDCs) extracted at the $\Gamma$ point (Fig.~\ref{Fig3}(b)), this hybridised state shifts away from the Fermi level with decreasing temperature. We monitor the temperature-dependent separation of this hybridised state from the top of the Te $5p$-derived valence bands as a fingerprint of the onset of long-range magnetic order here, where the cross-referencing to the valence band top ensures that temperature-dependent chemical potential shifts or residual sample charging due to the semiconducting bulk cannot affect our conclusions. From this (Fig.~\ref{Fig3}(c)), we find that the onset of the shift to higher binding energies of the hybridised state starts at around 120~K for the sample shown here with an electron doping of $N=3.1\times10^{13}$~cm$^{-2}$, considerably higher than for the undoped bulk where such energy shifts onset at the bulk $T_\mathrm{C}$ of $\approx65$~K. 

In fact, we find that the onset in energetic shifts of the Cr-Te hybridised state here is moved to further higher temperatures with increasing surface doping level (Fig.~\ref{Fig3}(c)). This indicates that the induced carriers are mediating an increased $T_\mathrm{C}$ here, consistent with both studies of the bulk where carriers are generated by organic ion intercalation with tetrabutyl ammonium~\cite{organic_CGT}, and with the observation of an enhanced $T_\mathrm{C}$ in electrostatically-doped CrGeTe$_3$~\cite{Verzhbitskiy_CGT}. To allow a quantitative comparison of the $T_\mathrm{C}$ enhancement, we estimate a baseline in our $E$ {\it{vs.}} $T$ curves (grey dashed line from Fig.~\ref{Fig3}(c)) and determine a $T_\mathrm{C}$ value as the temperature at which our measured band shifts cross a 10\% threshold (black dashed line from Fig.~\ref{Fig3}(c)). From this, we extract the doping-dependent phase diagram shown in Fig.~\ref{Fig3}(d). We observe a rapid increase in $T_\mathrm{C}$ from the bulk value already with our lowest electron doping. $T_\mathrm{C}$ then continues increasing with increasing doping density, also in line with the results from electrostatic gating which we include in our phase diagram~\cite{Verzhbitskiy_CGT}, but with a much slower rate. It thus appears that the introduction of free carriers in the relatively low-doping regime, at and around the point where the spin-minority states are populated and the van Hove singularity in the conduction band is traversed, are the most important for mediating a $T_\mathrm{C}$-enhancement in this system.

\begin{figure}[ht]
    \centering
    \includegraphics[width=\columnwidth]{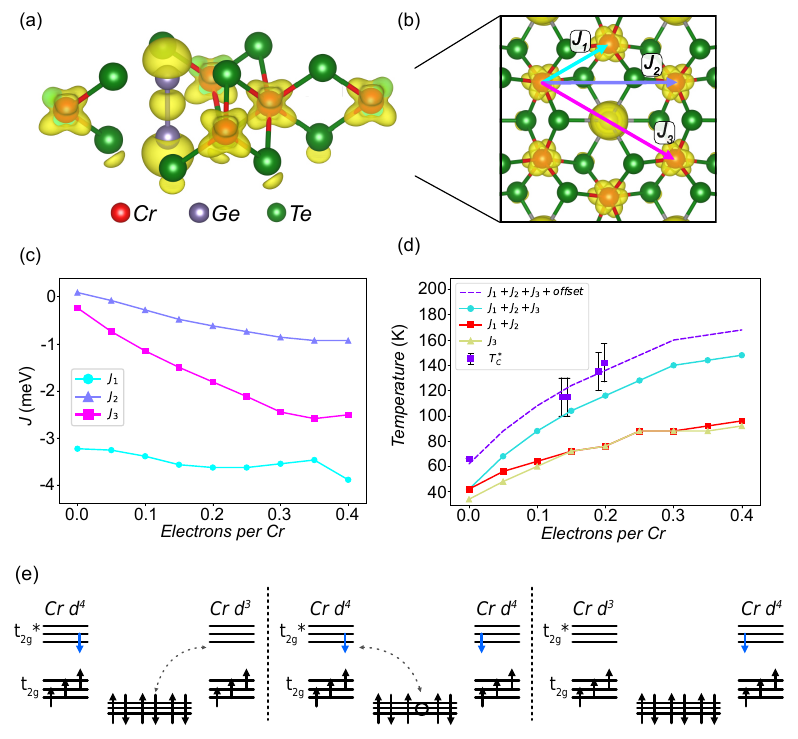}
    \caption{(a) Calculated charge density (yellow surface), integrated over an energy range of 0.2 eV around the spin-minority conduction band minimum and superimposed onto the crystal structure of CrGeTe$_3$ and (b) corresponding top-down view of the surface layer. The lobes of the charge density around each Cr site point in between the Cr-Te bond axis, a signature of $t_{2g}$ character, while the charge density around the Ge dimers are dramatically enhanced, pointing to a new conduction path involving the Ge sites. $J_1$, $J_2$, and $J_3$ paths are labeled with colour coded arrows. (c) Calculated values for $J_1$, $J_2$, and $J_3$, plotted as a function of doping concentration. The absolute value of $J_3$, in particular, shows a dramatic enhancement with doping. (d) $T_{\rm C}$ extracted from Monte Carlo simulations, considering different combinations of $J_{1,2,3}$, plotted alongside the experimental data. Here, $J_3$ is required (and has the equivalent impact of $J_1$ and $J_2$ combined) to match the shape of the experimental phase diagram.
    The dashed line through the experimental data points is the calculation for $J_1 + J_2 + J_3$ offset by 20~K, as a guide for the reader in showing the relative trends between experiment and simulation. (e) Schematic representation of the mechanism to enhance $T_\mathrm{C}$ enabled by the doped carriers in the minority-spin $t_{2g}$ manifold. Hopping paths from Cr to Te include direct Cr-Te-Cr hopping, and also effective Cr-Cr exchange pathways that are mediated via the Ge sites (i.e. $J_3$).}
    \label{Fig4}
\end{figure}

To explore the origins of this further, we study the evolution of the exchange parameters, $J_i$, as a function of doping from our DFT calculations, considering up to third-nearest neighbour terms ($i=3$, Fig.~\ref{Fig4}(a,b)). Interestingly, we find modest increases in both the nearest and next-nearest neighbour terms with increasing electron doping, and a rather dramatic increase in $J_3$ (Fig.~\ref{Fig4}(b)). We can understand this from the nature of the conduction band states which become populated upon electron doping. In an octahedral configuration, the naive choice for the next orbital to become occupied upon electron doping would be a spin-majority $e_g$ state.  Indeed, there is a significant amount of unoccupied $e_g$ spectral weight not far above the band gap, and virtual hopping processes into these states are thought to be relevant for the superexchange in the undoped system~\cite{Watson_CGT}. However, our calculated charge density distribution about the Cr sites for the lowest conduction band (Fig.~\ref{Fig4}(a)) exhibits lobes pointing between the neighbouring Te sites, indicating a $t_{2g}$-like character, consistent with the spin-minority nature of the states at the bottom of the conduction band. This band, however, sits well below the majority of the unoccupied $t_{2g}$ weight~\cite{kang2019effect}, and is surprisingly dispersive, dropping below the $e_g$ states at the K and M points. In fact, our calculations reveal a notable charge density of this band also around the Ge sites. This points to a significant hybridisation of the states at the bottom of the conduction band with Ge, which had typically been thought to play a relatively inert role in the magnetic properties of (undoped) CrGeTe$_3$. Considering the $J_3$ exchange pathway, however, (Fig.~\ref{Fig4}(a)), it is clear how in the electron-doped system, hopping via the Ge sites could mediate a significant increase in $J_3$.

We show in Fig.~\ref{Fig4}(c) an estimate from Monte Carlo calculations (see also Supplementary Fig.~S5) of the doping-dependent $T_\mathrm{C}$ that would be expected, given such a change in exchange parameters as calculated here. Considering the evolution of all calculated exchange constants (purple line in Fig.~\ref{Fig4}(c)), we find that the calculated $T_\mathrm{C}$ slightly underestimates the experimental values both for the bulk and the doped systems, but with a doping-dependent trend which closely follows our experimental results (see also offset calculation, purple dashed line in Fig.~\ref{Fig4}(c)). Excluding either the increase in $J_3$ (green line in Fig.~\ref{Fig4}(c)) or in $J_1$ and $J_2$ (red line in Fig.~\ref{Fig4}(c)), it is clear that approximately half of the increase observed here originates from the new hopping pathway which is opened up via the Ge dimers, shown in Fig.~\ref{Fig4}(a). 

\section{Discussion}
We attribute the resulting increase in $T_\mathrm{C}$ here to an unusual form of double exchange, mediated via a dispersive branch of the spin-minority $t_{2g}$ carriers, as summarised schematically in Fig.~\ref{Fig4}(e). In a simple picture, we can consider the undoped electronic structure of CrGeTe$_3$ as having a half-filled $t_{2g}$ manifold in a $d^3$ electron configuration~\footnote{We note that, in reality, even for the undoped compound, finite mixing with the ligand states plays a crucial role, ultimately underpinning the ferromagnetic order via a superexchange mechanism~\cite{Watson_CGT}.}. The doped carriers must therefore fill either the majority-spin $e_g$ states or the minority-spin $t_{2g}$ states. Our DFT calculations above indicate the latter occurs here at low doping, thus leading to an additional Cr spin on some sites which is anti-aligned with the majority spin moment. Given the dispersive nature of the $t_{2g}$-derived state at the bottom of the conduction band here, this doped carrier can readily delocalise through the crystal, hopping via the Te sites (including the Cr-Te-Ge-Te-Cr hopping pathway identified above). This, however, requires the creation of a hole on the Te site that has equal spin as the doped electron, which thus acts to mediate a net ferromagnetic coupling here (Fig.~\ref{Fig4}(e)). 

Since the doped carriers occupy the minority-spin $t_{2g}$ states, the net magnetic moment will be partially cancelled. Indeed, as shown in Supplementary Fig. S6, we find that the magnetic moment decreases linearly with the increase of doping concentration at low doping levels. As the doping concentration increases further, however, the spin-majority $e_g$ states start to become populated (Fig.~\ref{Fig2}(d)), resulting in fluctuations in the magnetic moment (Supplementary Fig. S6). Consistent with this, we find that the increase in $J$ and the associated increase in calculated $T_\mathrm{C}$ starts to saturate above a critical doping of around 0.3 electrons/Cr (Fig.~\ref{Fig4}(c,d)), in agreement with experiments (Fig.~\ref{Fig3}(d)). Our study thus highlights how the population of spin-minority conduction band states with electron doping in CrGeTe$_3$ mediates a more than doubling of its ferromagnetic $T_\mathrm{C}$, explaining recent findings of gate-tuned magnetism in this system. Our findings also highlight an unusual situation that occurs here under such electrostatic doping, with a half-metal forming at the surface which has opposite spin-polarisation to the bulk, thus suggesting potential for novel spintronic devices such as intrinsic, and gate-tunable, spin valve structures, magnetic tunnel junctions, or spin-filters~\cite{tang_2022}.

\section{Methods}
\noindent{\bf{Sample growth:}} CrGeTe$_3$ single crystals were prepared following the procedure in Ref.~\cite{Ji_CGT_Growth}. The crystals were separated from the flux by centrifuging at $500^{\circ}$C and then cleaned free of any excess Te. The crystals were found to exhibit a sharp magnetic transition at approximately 65 K~\cite{Watson_CGT}, in good agreement with previous literature reports for pristine bulk CrGeTe$_3$~\cite{Liu_Critical_behavior}. 

\

\noindent{\bf{Angle-resolved photoemission spectroscopy:}} ARPES measurements were performed at the high-resolution branch of the I05 beamline at Diamond Light Source, UK. Single crystal samples were cleaved in situ after being cooled to approximately 40~K, and then electron-doped by depositing Rb atoms on the surface via evaporation from a SAES getter source. To obtain variable electron density, the samples were dosed for varying time, between 30~s and 360~s. The chamber pressure did not exceed $7\times10^{-10}$~mbar during the deposition. 

ARPES measurements were performed using linearly polarised photons (with photon energies as specified in the figure captions) at sample temperatures ranging between 40 K and 200 K. Since CrGeTe$_3$ is a semiconductor, sample charging due to photoemission can become problematic at low temperatures~\cite{Baer_xps_charging}. To minimise the impact of this, we limited our measurement range to above 40 K and used rather low photon fluxes. To further rule out any influence of sample charging on our conclusions, we extract $T_\mathrm{C}$ based on analysis of energy differences in our measured spectra, as outlined in the text. 

The doping levels were estimated from our measured ARPES data using Luttinger's Theorem \cite{Luttinger_Theorem}, where the volume enclosed by the Fermi surface is proportional to the particle density by, 
\begin{equation*}
    N = g\int \frac{d^Dk}{(2\pi)^D},
\end{equation*}
where N is the number of charge carriers, D is the dimensionality of the system, and g is the spin and valley degeneracy, characterising how many equivalent bands exist within the first Brillouin zone, with $D=2$ here. The area enclosed by the Fermi surface was extracted from our fitted Fermi surface data by implementing a Green's Theorem algorithm. 

\

\noindent{\bf{Density functional theory and Monte Carlo calculations:}} Our calculations were carried out within the Perdew-Burke-Ernzerhof (PBE) generalized gradient approximation \cite{25perdew1996generalized}, as implemented in the Vienna $ab$ Initio Simulation Package (VASP) \cite{26kresse1993ab}. To better consider the vdW forces, we performed calculations with a non-local vdW density functional in the form of optB86b-vdW \cite{vdwklime2011van,vdwklimevs2009chemical}. The interactions of electrons with ionic cores were described using projector-augmented wave (PAW) pseudopotentials \cite{29blochl1994projector,30kresse1999ultrasoft}. For all the calculations, we chose the energy cutoff to be 600 eV, the maximum force of all atoms was optimized until it fell to less than 0.01~eV/{\AA}, and the criterion for the total energy convergence was set as $10^{-6}$ eV. The set of $12\times12\times1$ $k$-point samplings \cite{31monkhorst1976special}, in the reciprocal space by the Gamma-centered grid, was used to model the Brillouin zone. A vacuum space exceeding 20~{\AA} was included to avoid interaction between adjacent periodic layers for the monolayer calculations.

In this work, all of the geometry structures were fully relaxed, according to the maximum force and total energy convergence, listed previously. The lattice constant $a$ and $b$ of optimized monolayer CrGeTe$_3$ are 6.90~{\AA}, which is close to 6.91~{\AA} reported by Fang $et\ al$ \cite{lattice_fang2018large}. We calculated the energy of different dopant positions and found that the configuration in which the Rb is located directly above the Cr sites has the lowest energy. This superstructure was then used to calculate the band structure. Our calculations indicated that the effects of electron doping are dominated by a rigid shift of the band structure. For the calculation of the exchange parameters and magnetic moments as a function of electron doping, we thus artificially changed the number of electrons to simulate the non-integer doping concentration via a virtual crystal approximation. These calculations were performed without spin-orbit coupling. Whilst spin-orbit coupling leads to significant changes for the Te-derived valence bands~\cite{Watson_CGT}, we confirmed that they have minimal impact on the conduction band electronic structure, and on the calculated exchange parameters.

We used the VAMPIRE \cite{33evans2014atomistic} code to perform Monte Carlo (MC) simulations, assuming isotropic exchange constants up to the third nearest neighbour. Anisotropic effects are modelled via an effective single-ion anisotropy term, with coupling constant $A_i$ extracted from the magneto-crystalline anisotropy. $J$ was extracted by the energy mapping method~\cite{35xiang2013magnetic}, with a negative value indicating ferromagnetic interactions. The standard Metropolis algorithm~\cite{34asselin2010constrained} has been used for MC simulations with 5$\times$10$^4$ MC steps for equilibration and 3$\times$10$^5$ MC steps for averaging. Since the VAMPIRE software can only simulate cuboid cells, we used cuboid cells, which contain 4 Cr sites. We performed calculations for a $40\times40\times1$ supercell with $N_s$ = 6400 spins. The transition temperature can be estimated from the peaks that appear in the temperature evolution of specific heat (Supplementary Fig.~S5), evaluated as:
\begin{equation*}
    C_v = \frac{k_B \beta ^2}{N_s} [\langle E^2\rangle-\langle E\rangle^2]
\end{equation*}
where $E$ is the energy calculated using VAMPIRE, $k_B$ is the Boltzmann constant and $\beta$ = $1/k_BT$, while $\langle ...\rangle$ indicates statistical averages.

\section{Acknowledgements}
We thank Philip Murgatroyd, Bernd Braunecker, and Chris Hooley for useful discussions. We gratefully acknowledge support from The Leverhulme Trust via Grant No.~RL-2016-006, and the UK Engineering and Physical Sciences Research Council via Grant Nos.~EP/X015556/1, EP/T005963/1, and EP/N032128/1. P.K. and S.P. acknowledge support from the Royal Society through the International Exchange grant IEC\textbackslash{}R2\textbackslash{}222041. IM and EAM gratefully acknowledge studentship support from the International Max-Planck Research School for Chemistry and Physics of Quantum Materials. We thank Diamond Light Source for access to Beamline I05 (Proposals SI21986 and SI25564), which contributed to the results presented here. S.P. acknowledges support from the Ministry of University and Research (MUR) PRIN 2022 project ``SORBET - Spin ORBit Effects in Two-dimensional magnets", project No. 2022ZY8HJY.  L.Q. acknowledges the support from China Scholarship Council, W.R. acknowledges the support from the National Natural Science Foundation of China (12074241, 11929401, 52120204), and the Science and Technology Commission of Shanghai Municipality (22XD1400900, 20501130600). For the purpose of open access, the authors have applied a Creative Commons Attribution (CC BY) licence to any Author Accepted Manuscript version arising. The research data supporting this publication can be accessed at [[DOI TO BE INSERTED]].

\addcontentsline{toc}{section}{References}
\bibliographystyle{my_bst.bst}

\newpage 

\renewcommand{\figurename}{Supplementary Fig.}
\setcounter{figure}{0}   

\begin{figure*}[!h]
    \centering
    \includegraphics[width=\textwidth]{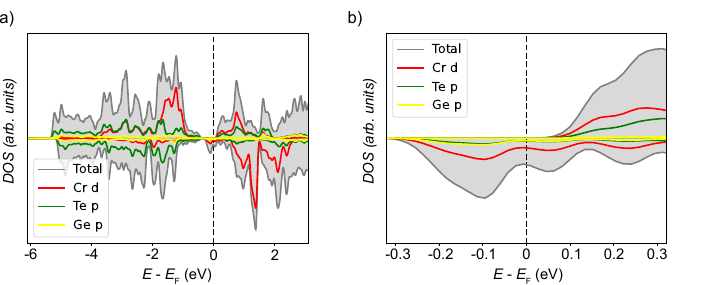}
    \caption{Calculated spin-resolved partial density of states for electron-doped (N = 0.2 electrons per Cr) monolayer CrGeTe$_3$, for the spin-majority (positive y-axis) and minority (negative y-axis) states. (a) Orbital character of the valence band. (b) Orbital character in an energy range corresponding to the partially filled conduction band.}
    \label{fig:DOS}
\end{figure*}

\begin{figure*}[h]
    \centering
    \includegraphics[width=0.4\textwidth]{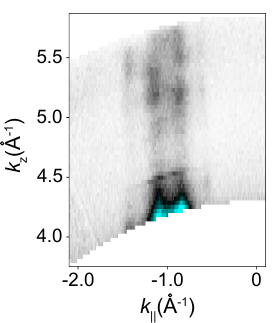}
    \caption{Out-of-plane dispersion for electron-doped CrGeTe$_3$ at $E=E_{\rm F}$. The conduction bands lack dispersion in the out-of-plane direction, indicating two-dimensionality.}
    \label{fig:hubbard}
\end{figure*}

\begin{figure*}[h]
    \centering
    \includegraphics[width=\textwidth]{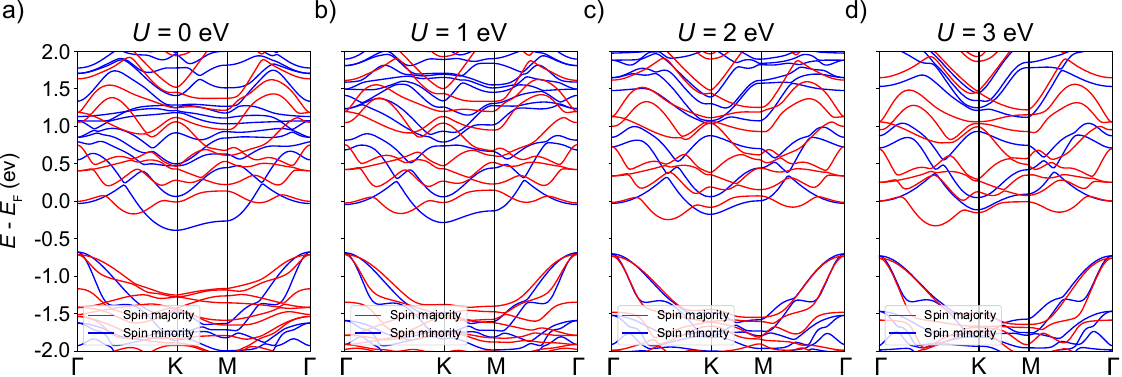}
    \caption{Electronic band structure of electron-doped (N = 0.2 electrons per Cr) monolayer CrGeTe$_3$, calculated as a function of Hubbard parameter $U$. Increasing $U$ drives an inversion between the spin-minority and spin-majority states at the bottom of the conduction band. Only the low-$U$ limit ($U<2$~eV) is consistent with our experimental measurements.}
    \label{fig:hubbard}
\end{figure*}

\begin{figure*}[h]
    \centering
    \includegraphics[width=0.8\textwidth]{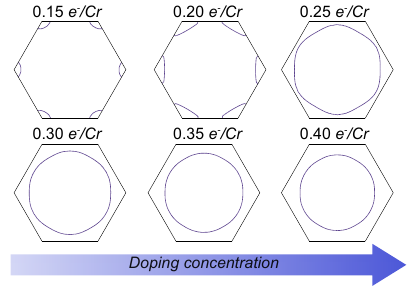}
    \caption{Calculated Fermi surface of electron-doped CrGeTe$_3$ as a function of electron doping, revealing the topological Lifshitz transition of the Fermi surface with increasing electron doping.}
    \label{fig:hubbard}
\end{figure*}

\begin{figure*}[h]
    \centering
    \includegraphics[width=\textwidth]{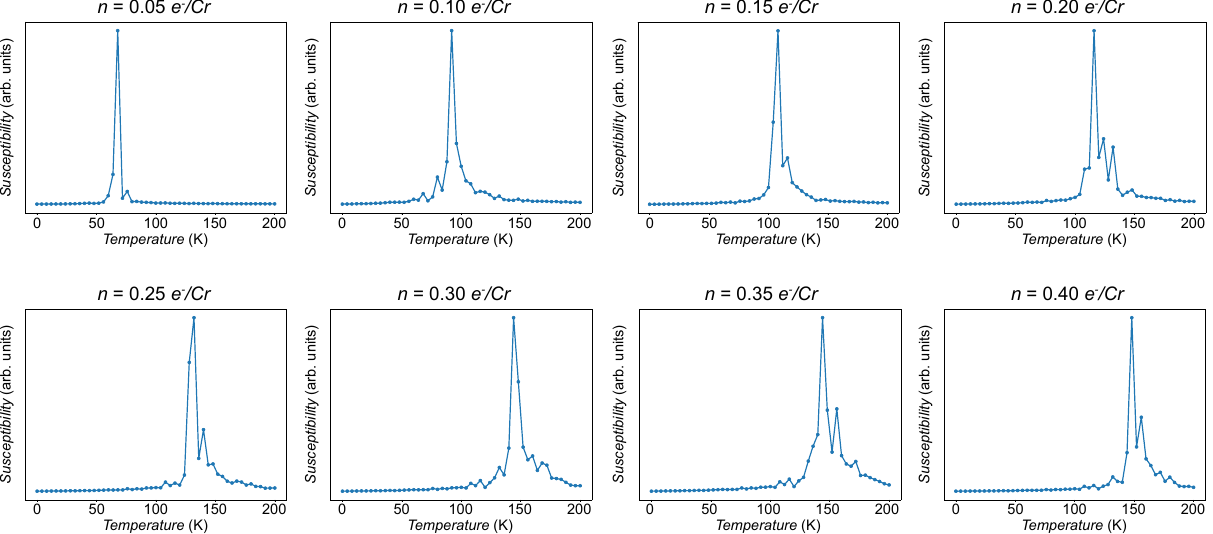}
    \caption{Magnetic susceptibilities calculated, as a function of electron doping, from Monte Carlo simulations. Transition temperatures are estimated from the location of discontinuities in the susceptibility.}
    \label{fig:magmoml}
\end{figure*}

\begin{figure*}[h]
    \centering
    \includegraphics[width=0.6\textwidth]{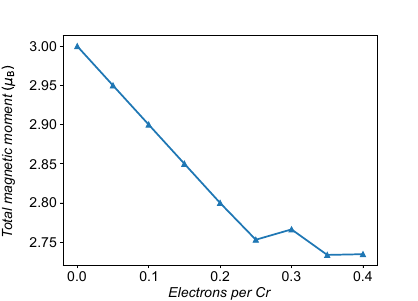}
    \caption{Calculated total magnetic moment, as a function of electron doping. At doping levels above $0.25~e^-/Cr$, the spin-majority Cr $e_g$ derived conduction band starts to be populated, halting the monotonic decrease in magnetic moment with increasing carrier number evident at lower doping levels.}
    \label{fig:magmoml}
\end{figure*}

\end{document}